\documentclass[12pt]{article}
\usepackage{epsfig,rotating}
\usepackage{cite}
\usepackage{xspace}
\def\XVEC#1#2{#1_1,\ldots,{#1}_{#2}}  
\def\hs{\hskip}

\def\aver#1{\left\langle#1\right\rangle}  
\newcommand{\rd}{{\mbox{d}}}

\newcommand{\beq}{\begin{equation}}
\newcommand{\eeq}{\end{equation}  }

\newcommand{\bec}{\begin{center}}
\newcommand{\eec}{\end{center}}

\newcommand{\text}[1]{\mbox{\rm #1}}

\def\eg{{\sl e.g.\xspace\/}}  

\newcommand{\Wp}{\mbox{w}^+}
\newcommand{\Wn}{\mbox{w}^-}
\newcommand{\W}{\mbox{w}}
\newcommand{\WW}{\mbox{ww}}
\newcommand{\rhomix}{\rho_2^{\mbox{\rm\small mix}}}

\newcommand{\WWqqqq}{\mathrm{W}^+\mathrm{W}^-\to
\bar{\mathrm{q}}_1\bar{\mathrm{q}}_2\mathrm{q}_3\mathrm{q}_4}

\begin{document}
\clearpage
\pagestyle{empty}
\setcounter{footnote}{0}\setcounter{page}{0}%
\thispagestyle{empty}\pagestyle{plain}\pagenumbering{arabic}%

\hfill {}

\hfill {}

\hfill {}

\vspace{1.0cm}

\begin{center}

\vskip 0.8in plus 2in

{\Large\bf Correlations in \boldmath{$e^+e^-\to W^+W^-$} hadronic decays 

\vspace{2.0cm}

{\rm\large E.A.~De~Wolf\\[0.5cm]  
{\noindent\rm \small\em  EP Division, CERN,  European Organisation for Nuclear Research,\\
CH-1211 Geneva 23, Switzerland\\ and\\  
Department of Physics, Universitaire Instelling Antwerpen, \\ Universiteitsplein 1, B-2610 Antwerpen, Belgium.
}}
}
\normalsize
\date{}

\vspace{2.0cm}

\begin{abstract}
\noindent A mathematical formalism for the analysis of correlations in
multi-source events such as $W^+W^-$ production in $e^+e^-$ annihilations
is presented. Various   measures  used in  experimental
searches for inter-$W$  correlations are  reviewed.
\end{abstract}

\end{center}
\newpage
\setcounter{page}{1}
\section{Introduction}
%
The problem of possible inter-$W$ dynamical correlations  continues to
attract much experimental attention at LEP-II~\cite{kittel:review}. 
The importance  of these in a precision
measurement of the $W$-mass~\cite{sjo,kh,t1} is by far the main reason 
for the flurry of present activities. 
However, this  should not be the sole motivation for  careful
experimental work in this subfield. For example,  the possible absence  
of Bose-Einstein correlations (BEC) between pions originating from different $W$'s 
raises interesting general and basic questions 
regarding the coherent versus incoherent nature of
particle emission.  Also the question of possible color- or string-reconnection
effects may be of considerable importance for our  understanding of the vacuum properties of 
Quantumchromodynamics.

As repeatedly emphasized by the Lund group~\cite{bo:book,bo:ringner}, 
Bose-Einstein correlations
of a {\em coherent} type, for which we suggest the name 
  ``String Symmetrization Correlations (SSC)'',
are present in any string model of hadron production. Moreover, the SSC depend 
essentially only  on {\em local\/} properties of the string and should thus be
 independent of the
 environment in which the string is fragmenting. 
This basic property is not in contradiction with  
experimental results on Bose-Einstein correlations in $e^+e^-$ annihilations and
lepton-nucleon scattering.

However, in systems comprising  several strings, a {\em second type\/} of Bose-Einstein
correlations may exist if the different strings behave independently so that they act
as incoherent sources of particle emission.
This second-order intensity correlation effect, or HBT effect, since first discovered by
 Hanbury Brown and Twiss~\cite{hbt},  is expected to reflect the ``geometry'' of the
collision process and, in particular, to  be sensitive to the size of the ``freeze-out'' 
volume where the hadrons are formed. This volume could have an extension of many fermi's and 
can thus be significantly larger than the typical ``radii'' of less than a fermi,
commonly measured {\eg} in BEC studies of $e^+e^-$ annihilations. 
As a result,  correlation functions of identical bosons
may well  show an additional enhancement at substantially smaller values of their
momentum difference than for SSC correlations.

The reaction $e^+e^-\to \WWqqqq\to\mbox{hadrons}$ 
is a prototypical example
of a system of which the hadronisation  proceeds via  the fragmentation of
two color-fields or strings. Barring possible reconnection effects at an early stage of
the evolution of the system, these strings are thought to fragment independently.
Thus, SSC within single $W$'s and  inter-$W$ HBT correlations could  coexist in this system, but
manifest themselves  in different ranges of {\eg} the commonly used variable $Q^2$, 
the square of the difference of the four-momenta of the identical pions. 
No dedicated searches in this direction have been made so far. Moreover, if the
HBT correlations are of much shorter range than naively expected, the influence on
$W$-mass measurements is likely to be much weaker than present studies suggest. 

Limited statistics as well as limited experimental sensitivity at 
very small $Q$ may  prevent
the HBT-type of correlation effect to be clearly observed in  $e^+e^-\to W^+W^-\to 4q$
events. However,  as yet unexplored alternative reactions exist.
 Much larger  statistics is available in {\eg} $e^+e^-\to q\bar{q} +g \to\mbox{\rm three-jet}$ events.
For  these, due to the emission of a hard gluon, two color-disconnected
strings are stretched, one  between the quark and the gluon, another between the 
gluon and the antiquark.  Depending on the origin of the identical pions studied,
SSC as well as HBT correlations should contribute\footnote{I thank B.~Andersson and G. Gustafson
for discussions of this point.} to  the correlations among identical bosons.

In hadron-hadron, hadron-nucleus and nucleus-nucleus collisions, hadroproduction is also
believed to result from  the break-up of several to very many strings.
Since the superposition of many independent particle ``sources'' 
weakens the measured strength of  correlations of the SSC type,
HBT correlations may well dominate
the  Bose-Einstein correlations  measured in these processes. This could also  explain the
observed positive correlation between particle density (or multiplicity) and the measured 
BEC radii. Finally, due to the ``inside-outside'' character of
hadron production, whereby low-momentum particles ``freeze out'' first (in any reference frame), 
one may expect a correlation between the width of the Bose-Einstein enhancement
and particle momenta if the HBT correlations  are the dominant effect.

Returning to correlations in the $WW$ system,
 experimental study of such effects should start from a well-defined mathematical framework.
Since, in general, inter-$W$ dynamics introduces genuine correlations between the
decay products of the $W$'s, we 
generalize the formalism  presented 
in an earlier paper on the subject~\cite{edw:beww} where observables relevant
for the case of stochastic independence and 
fully overlapping decays of the $W$'s were proposed. We  
consider the general case of  stochastic  dependence and the separation of the $W$ hadronic decay
products in momentum space. 
We concentrate on second-order (two-particle) inclusive densities and correlation functions 
but the results can be  generalized to higher orders. 

In the mathematical treatment of the problem, a general inter-$W$ correlation function is introduced,
representing an arbitrary stochastic correlation between hadrons from different $W$'s which could arise
from color reconnection effects, Bose-Einstein correlations or others.
Although most of the examples treated  in this paper relate to Bose-Einstein studies involving
identical particles,
the formalism is general and can be used, where needed 
with a suitable change of kinematic variables, in other contexts as well.

\section{Multivariate distributions and moments\label{sec:multiv}}
Before discussing the problem of correlations among particles originating
from   $W$'s decaying into  fully hadronic, so-called four-jet configurations,
$e^+e^-\to q_1\overline{q}_2 q_3 \overline{q}_4\to\mbox{hadrons}$, 
we first consider  a more general problem.

Assume that a system ({\eg} an event) comprising in total $n$ (observed) particles,   
can be subdivided into $S$ possibly stochastically correlated  
groups or ``sources'' $\Omega_i$ ($i=1,\ldots,S$). 
We take the $n$ particles to be of identical  type and assume
that the groups  are mutually exclusive i.e. every particle can be assigned  
to one and one group only\footnote{Such a subdivision can be based on
a ``natural'' partition of the $n$ particles in $S$ groups according to the
underlying dynamics. It could  also be based on an experimentally dictated partition
of the phase space in $S$ distinct (non-overlapping) regions {\eg}  as the result of
jet clustering.}: 
 $\Omega_i\cap\Omega_j=0$, $\forall i,j$ and $i\ne j$. No assumptions are made, however,
about possible overlap in momentum space of  particles from different groups.

Let $n$ be the total number of particles counted in the union of the $S$ groups,
\begin{equation}
n=\sum_{m=1}^Sn_m\ \ .
\label{dr2:2}
\end{equation}
Consider further the multivariate multiplicity distribution
$P_S(\XVEC{n}{S})$ giving the joint probability for the simultaneous
occurrence of $n_1$ particles in  class $\Omega_1$, $\ldots,$ $n_S$
particles in class $\Omega_S$. 
The probability distribution of $n$ is then given by
\begin{equation}
P(n)=\sum_{n_1=0}^{n}\cdots\sum_{n_S=0}^{n} P_S(\XVEC{n}{S})
\delta_{n,n_1+\cdots+n_S}\, .
\label{dr2:3}
\end{equation}
We further  define  the single-variate factorial moment generating function of $P(n)$
\begin{equation}
G(z)=\sum_{n=0}^\infty(1+z)^n\,P(n)\,.
\label{dr2:4}
\end{equation}
By expanding $(1+z)^n$,  (\ref{dr2:4}) can be rewritten as
\begin{eqnarray}
G(z) &=&1+\sum_{q=1}^\infty \frac{z^q}{q!} \sum_{n=0}^\infty \frac{n!}{(n-q)!}\, P(n),\label{xx1}\\
     &=&1+\sum_{q=1}^\infty \frac{z^q}{q!} \, \aver{ \frac{n!}{(n-q)!} },\label{xx2}\\
G(z) &=&1+\sum_{q=1}^\infty \frac{z^q}{q!} \,\tilde{F}_q;\label{dr2:4b}
\end{eqnarray}
where in the last line we introduce
 the symbol $\tilde{F}_q$ for the (unnormalized)  
factorial (or binomial) moment of the  probability distribution $P(n)$. The brackets in Eq.~(\ref{xx2})
denote a statistical average.
\begin{eqnarray} 
 \tilde{F}_q &=& \aver{n(n-1)\ldots(n-q+1)},\nonumber\\
 &=& \int_\Delta \rd y_1\ldots\int_\Delta \rd y_q\;
\rho_q(\XVEC{y}{q}).\label{eq:int}
\end{eqnarray} 
The last equation expresses that, for identical particles,
the factorial moment $\tilde{F}_q$ is equal to the integral over
$q$-dimensional phase space (here  for simplicity of notation represented by the variables $y_i$) 
of the $q$-particle inclusive density $\rho_q(\XVEC{y}{q})$ over the same phase space volume
$\Delta$~\cite{edw:review}.
The    $q$-particle inclusive density $\rho_q(\XVEC{y}{q})$ is defined as
\begin{equation}
 \rho_q(\XVEC{y}{q})=\frac{1}{\sigma}\,\frac{d^q\sigma}{dy_1\ldots dy_q},
\label{eq:def:rhoq} 
\end{equation}
with $\sigma$ the total cross section of the considered reaction.
Experimentally, this quantity is approximated by
\begin{equation}
 \rho_q(\XVEC{y}{q})=\frac{1}{N_{\mbox{\rm\small evt}}}
 \,\frac{dN^{\mbox{\rm\small q-tuples}}}{dy_1\ldots dy_q},
\label{eq:def:rhoq:expt} 
\end{equation}
with $dN^{\mbox{\rm\small q-tuples}}$ the number of $q$-tuples of particles, counted in
a phase space domain $(y_1+dy_1,\ldots,y_q+dy_q)$; $N_{\mbox{\rm\small evt}}$ is the number
of events in the sample.

Factorial cumulants are formally defined as the coefficients of ${z^q}/{q!}$ in the
Taylor expansion of the function $\log{G(z)}$:
\begin{equation}
\log G(z)=\aver{n}z +\sum_{q=2}^{\infty} \frac{z^q}{q!}\; \tilde{K}_q.
\label{dr:17}
\end{equation}

The (unnormalized) factorial cumulants  of order $q$, $\tilde{K}_q$,  also known as Mueller
moments~\cite{Mue71}, are  equal to the  $q$-fold phase space integral
of the $q$-particle inclusive (so-called ``connected'' or ``genuine'') factorial cumulant
 correlation function  $C_q(\XVEC{y}{q})$ 
\begin{equation}
\tilde{K}_q=\int_\Delta \rd y_1\ldots\int_\Delta \rd y_q\, C_q(\XVEC{y}{q}). 
\label{dr:18}
\end{equation}

The correlation functions  $C_q(\XVEC{y}{q})$ are, as in the cluster expansion in statistical
mechanics, defined via the sequence
\begin{eqnarray}
\rho_1(1)& =& C_1(1),\nonumber\\
\rho_2(1,2)& =& C_1(1)C_1(2) +C_2(1,2),\nonumber\\
\rho_3(1,2,3)& =& C_1(1)C_1(2)C_1(3)
+C_1(1)C_2(2,3)
+C_1(2)C_2(1,3)
+\nonumber \nonumber \\
& &\mbox{}
+C_1(3)C_2(1,2)+C_3(1,2,3),\\
\mbox{etc.}
\end{eqnarray}
These relations  can be inverted yielding
\begin{eqnarray}
C_2(1,2)&=&\rho_2(1,2) -\rho_1(1)\rho_1(2)\ ,\nonumber\\
C_3(1,2,3)&=&\rho_3(1,2,3)
-\sum_{(3)}\rho_1(1)\rho_2(2,3)+2\rho_1(1)\rho_1(2)\rho_1(3)\ ,\nonumber\\
C_4(1,2,3,4)&=&\rho_4(1,2,3,4)
-\sum_{(4)}\rho_1(1)\rho_3(1,2,3)
-\sum_{(3)}\rho_2(1,2)\rho_2(3,4)\nonumber\\
&&\mbox{} +2\sum_{(6)}\rho_1(1)\rho_1(2)\rho_2(3,4)-6\rho_1(1)
\rho_1(2)\rho_1(3)\rho_1(4),\\
\mbox{etc.}
\label{a:4b}
\end{eqnarray}
In the above relations  we have abbreviated $\rho_q(\XVEC{y}{q})$ to
 $\rho_q(1,2,\ldots,q)$ etc.; the summations indicate that all possible permutations
have to be taken (the number under the summation sign indicates the number of  terms).

The factorial moments $\tilde{F}_q$ and factorial cumulants $\tilde{K}_q$ are easily found if $G(z)$ is
known
\begin{eqnarray}
\tilde{F}_q&=& \left.\frac{\rd^q G(z)}{\rd z^q}\right|_{z=0}\ \ \ ,\\
\tilde{K}_q        &=& \left.\frac{\rd^q \log{G(z)}}{\rd z^q}\right|_{z=0}.
\end{eqnarray}
The counting distribution $P(n)$ is likewise determined by $G(z)$
\begin{equation}
P(n)= \frac{1}{n!}\left.\frac{\rd^n G(z)}{\rd z^n}\right|_{z=-1}.
\end{equation}
Let us now introduce the  multidimensional $S$-variate generating function   
\begin{eqnarray}
\hs-7mm G_S(\XVEC{z}{S})&=&\sum_{n_1=0}^\infty\sum_{n_2=0}^\infty\cdots
\sum_{n_S=0}^\infty
\;(1+z_1)^{n_1}\cdots(1+z_S)^{n_S}\,P_S(\XVEC{n}{S}),\label{dr:51}
\end{eqnarray}
from which the $S$-variate factorial moments are easily obtained by differentiation:
\begin{eqnarray}
\hs-7mm \tilde{F}_{q_1\dots q_S}=\aver{n_1^{[q_1]}\dots n_S^{[q_S]}}&=&
\left. \left( \frac{\partial}{\partial z_1}\right)^{q_1}\cdots\left(
\frac{\partial}{\partial z_S}\right)^{q_S}\; G_S(\XVEC{z}{S})
\right|_{z_1=\cdots z_S=0} . \label{dr:53}
\end{eqnarray}
Likewise,  $S$-variate factorial cumulants are  obtained by differentiation of 
$\log{G_S(\XVEC{z}{S})}$. 

Returning to the function $G(z)$ (\ref{dr2:4}), it is not difficult to see that it 
can be written in terms of the multivariate generating function (\ref{dr:51}) as
\begin{equation}
G(z)=\left.G_S(\XVEC{z}{S})\right|_{z_1=z_2=\cdots=z_S=z}.  \label{dr2:5} 
\end{equation}
Equation~(\ref{dr2:5}) therefore allows to express the factorial moments of $n$ in terms of the
multivariate factorial moments of $\{\XVEC{n}{S}\}$.
  
Application of the Leibnitz rule 
$$
\left(\frac{\rd}{\rd z}\right)^q f(z)=\sum_{\{a_j\}}\frac{q!} {a_1\,!a_2\,!\ldots a_S!}
\left(\frac{\rd}{\rd z}\right)^{a_1}\,f_1(z)\cdots
\left(\frac{\rd}{\rd z}\right)^{a_S}\,f_S(z) 
 $$ 
to the function 
$$f(z)=f_1(z)\cdots f_S(z)$$ 
and using (\ref{dr2:5}) leads immediately to the relation 
\begin{equation} 
\tilde{F}_q=\sum_{\{a_j\}}
\tilde{F}^{(S)}_{a_1\ldots a_S}\, \frac{q!}%
{a_1!\,\ldots\,a_S!}.  \label{dr2:6}
\end{equation} 

The summation is over all sets $\{a_j\}$ of non-negative integers such that
$$\sum_{j=1}^{S}a_j=q.$$ 
Formula (\ref{dr2:6}) is  a generalization for factorial moments of the
usual multinomial theorem. 

  Likewise, taking the natural logarithm of both sides
of (\ref{dr2:5}), one obtains an identical relation as (\ref{dr2:6}) among single-variate
and multivariate factorial cumulants.

As an example, for two groups ($S=2$) one finds from  (\ref{dr2:6}) 
\begin{eqnarray}
\tilde{F}_2 &=& \tilde{F}^{(2)}_{02}
+2\tilde{F}^{(2)}_{11}+
\tilde{F}^{(2)}_{20}\ ,\label{eq:start1}\\
\tilde{F}_3 &=& \tilde{F}^{(2)}_{03}
+3(\tilde{F}^{(2)}_{12}+\tilde{F}^{(2)}_{21}) +
\tilde{F}^{(2)}_{30}\ ,\label{fff} \\
\tilde{F}_4 &=& \tilde{F}^{(2)}_{04}+6\tilde{F}^{(2)}_{22}
+4(\tilde{F}^{(2)}_{13}+\tilde{F}^{(2)}_{31})
+\tilde{F}^{(2)}_{40}\,.\label{eq:start2}
\end{eqnarray}

The factorial moments $\tilde{F}_{0i}$, $\tilde{F}_{i0}$,    are
determined  from the counting distribution in a single group. The 
 univariate factorial moments $\tilde{F}_q$ are obtained from the
sum of counts in the two groups. The ``mixed'' factorial moments $\tilde{F}^{(2)}_{ij}$ ($i,j\neq0$)
express inter-group stochastic dependences.
The relations (\ref{eq:start1})-(\ref{eq:start2}) are trivially  extended
to  more than two groups.

\section{\boldmath{$W^+W^-$} correlations}
\subsection{Integral moments and cumulants\label{sec:int}}
We now   apply  the general results from the previous section 
to the case of interest: the production of $W^+W^-$ in $e^+e^-$ annihilation, and their
subsequent decay  to four jets: $W^+W^-\rightarrow 4q\rightarrow \mbox{\rm \ hadrons}$.
Here, the number of ``sources'' $S$ is equal to two.

Eq.~(\ref{eq:start1}) is of particular interest. In a less formal way,
it can be written as:
\begin{equation}
\tilde{F}_2 \equiv\aver{n(n-1)}=
\aver{n_1(n_1-1)}+ \aver{n_2(n_2-1)} +2\aver{n_1n_2},
\,\label{eq:start11}\\
\end{equation}
where $n_1$ and $n_2$ are the number of particles from the decay of $W^+$ and
$W^-$, respectively.
 This equation could also be derived 
directly by noting that $n=n_1+n_2$ and working out the expression for $\aver{n(n-1)}$.
To derive  relations for $S>2$ or between  factorial moments of higher order, 
it is evidently less cumbersome to make use of the generating functions and
Eq.(\ref{dr2:6}).

In absence of inter-$W$ correlations {\em of whatever origin\/}, kinematical, 
dynamical, due to experimental selections and cuts, \dots,
one  has 

\begin{equation}
\aver{n_1n_2}=\aver{n_1}\aver{n_2}.
\,\label{eq:start12}
\end{equation}
Exhibiting explicitly the presence of  inter-$W$ correlations 
 we write (\ref{eq:start11}) as
\begin{equation}
\tilde{F}_2 \equiv\aver{n(n-1)}=
\aver{n_1(n_1-1)}+ \aver{n_2(n_2-1)} +2\aver{n_1}\aver{n_2}\,(1+\delta_I),
\label{eq:start13a}
\end{equation}
with 
\begin{equation}
\delta_{I}\equiv\aver{n_1n_2}/\aver{n_1}\aver{n_2}-1\ne0, \end{equation}
a measure  of positive or negative inter-$W$ correlations.
Similarly, the factorial cumulant $\tilde{K}_2$ can be written as 
\begin{equation}
\tilde{K}_2 =\tilde{K}_{20} + 
\tilde{K}_{02} +2\aver{n_1}\aver{n_2}\,\delta_I.\label{eq:start13b}
\end{equation}
This equation expresses the  correlation function, integrated over full phase space,
 of the whole system in terms of  
the  integrated correlation functions of each component separately, and of the
integrated inter-$W$ correlation. 
\subsubsection{Interlude\label{sec:interlude}}
The quantity  $\delta_{I}$ is related to the variances $D^2_{W^\pm}$ of the
single-$W^{\pm}$, $W^\pm\to q\overline{q}$, and $W^+ W^-\to 4q$ multiplicity distributions via the relation 
\begin{eqnarray}
D^2_{WW}&=&\aver{(n-\aver{n})^2}=\aver{[(n_{1}-\aver{n_{1}})+ (n_{2}-\aver{n_{2}})]^2}\\
&=&D^2_{W^+}+D^2_{W^-} +2\, \aver{n_{1}n_{2}} -2\aver{n_{1}}\aver{n_{2}}
\label{eq:start14} 
\end{eqnarray}
giving
\begin{equation}
\delta_I=
\frac{D^2_{WW}-D^2_{W^+}-D^2_{W^-}}{2\aver{n_{1}}\aver{n_{2}}}.
\label{eq:start15}
\end{equation} 

\subsection{Differential distributions\label{sec:differential}}
In section~\ref{sec:int},  general relations were obtained  relating 
factorial moments and cumulants of the multiplicity distributions of
 a $W^+W^-$ system      to  those of its  $W^+$ and $W^-$ components.      
Since  $q$-th order factorial moments
are integrals over phase space of $q$-particle inclusive 
densities, we now turn to relations among the fully differential particle densities
and correlation functions of  a $W^+W^-$ system 
and those of its composing parts, considering explicitly possible 
statistical  dependences. We restrict 
the discussion to second-order densities and correlations.

For two   stochastically independent systems, we derived in an earlier paper~\cite{edw:beww} 
the  relations
\begin{eqnarray}
C_2^{\WW}(1,2)  &=& C_2^{\Wp }(1,2) + C_2^{\Wn }(1,2)\label{r1},\label{k10} \\
\rho_2^{\WW}(1,2)&=&\rho_2^{\Wp}(1,2) + \rho_2^{\Wn}(1,2)
 +\rho_1^{\Wp}(1)\rho_1^{\Wn}(2)  +\rho_1^{\Wp}(2)\rho_1^{\Wn}(1),\label{k11} \\
\noalign{and further}
\rho_1^{\WW}(1)&=&\rho_1^{\Wp} (1) + \rho_1^{\Wn} (1).\label{r1x}
\end{eqnarray}
Here $C_2^{\WW}(1,2)$ and $C_2^{W^{\pm}}(1,2)$ are 
the two-particle correlation functions 
for $W^+W^-\rightarrow 4q$ events and  
$W^{\pm}\rightarrow 2q$ events, respectively;
$\rho_1^{\WW}(1)$,  $\rho_1^{\mbox{w}}(1)$ are the corresponding single-particle inclusive
densities. 

Inspection of  (\ref{eq:start13b}) suggest to write a {\em general} expression for
 $C_2^{\WW}(1,2)$ as
\begin{equation} 
C_2^{\WW}(1,2)=C_2^{\Wp }(1,2) + C_2^{\Wn }(1,2) + \delta_I(1,2)
\left\{\rho_1^{\Wp}(1)\,\rho_1^{\Wn}(2) + \rho_1^{\Wp}(2)\,\rho_1^{\Wn}(1)\right\}.
 \label{eq:cumulww} 
\end{equation}
The  function
$\delta_I(1,2)$ describes   correlations among different $W$'s. 

For  $\delta_I(1,2)=0$ (independent $W$'s), (\ref{eq:cumulww})
expresses the additivity\footnote{For  independent ``sources'', 
additivity is valid for all orders of the cumulant correlation functions.}
of the factorial cumulant correlation functions, a
necessary and sufficient condition for stochastic independence\cite{edw:beww}.  %

The factors $ \rho_1^{\Wp}(1)\,\rho_1^{\Wn}(2)$, $\rho_1^{\Wp}(2)\,\rho_1^{\Wn}(1)$ 
are introduced for normalization and explicitly 
account for differences in the single-particle
densities of the two $W$'s, as is the case when the momentum spectra of the
 decay products from different $W$'s are not identical i.e. do not fully overlap%
\footnote{In~\cite{edw:beww} it was implicitly assumed that
$
\rho_1^{\Wp}(1)\,\rho_1^{\Wn}(2)=\rho_1^{\Wp}(2)\,
\rho_1^{\Wn}(1)\equiv\rho_1^{\W}(1)\,\rho_1^{\W}(2),
\label{eq:overlap:zero} 
$ 
where $\rho_1^{\W}(1)$ is the single-particle inclusive density of one $W$ 
and with the  further  assumption 
that $\rho_1^{\Wp}(1)=\rho_1^{\Wn}(1)\equiv\rho_1^{\W}(1)$. Ignoring possible charge-dependence,
these equations are valid if the $W^+$ and $W^-$ hadronic decay products overlap completely
in momentum space.}.

With the definitions 
\begin{eqnarray} 
\rho_2^{\WW}(1,2)&=&C_2^{\WW}(1,2)+ \rho_1^{\WW}(1)\,\rho_1^{\WW}(2)\label{eq:define:rho2}\\
\rho_2^{\W}(1,2) &=&C_2^{\W}(1,2)+ \rho_1^{\W}(1)\,\rho_1^{\W}(2)
  \label{eq:define:rho1} 
\end{eqnarray} and using (\ref{r1x}) we can write a general form for $\rho_2^{\WW}(1,2)$ 
\begin{eqnarray} 
\rho_2^{\WW}(1,2)&=& C_2^{\Wp}(1,2) + C_2^{\Wn}(1,2) 
+ \left\{\rho_1^{\Wp}(1)\, \rho_1^{\Wn}(2) + 
\rho_1^{\Wp}(2)\, \rho_1^{\Wn}(1)\right\}\delta_I(1,2)\nonumber\\
&+& \left\{\rho_1^{\Wp}(1)+\rho_1^{\Wn}(1)\right\}
\,  \left\{\rho_1^{\Wp}(2)+\rho_1^{\Wn}(2) \right\}.
\label{eq:rho2b} 
\end{eqnarray}
Note that, in general, and in fully differential form, the terms 
$\rho_1^{\Wp}(1)\, \rho_1^{\Wn}(2)$ and  
$\rho_1^{\Wp}(2)\, \rho_1^{\Wn}(1)$ are {\em not\/} equal. $\rho_2^{\WW}(1,2)$, however,
 must be symmetric in its arguments for identical particles.
Equation  (\ref{eq:rho2b}), together with  (\ref{eq:define:rho1}) takes the form 
\begin{eqnarray}
\rho_2^{\WW}(1,2)&=& \rho_2^{\Wp}(1,2) + \rho_2^{\Wn}(1,2)\nonumber\\ 
&+&\left\{1+\delta_I(1,2)\right\}\, \left\{\rho_1^{\Wp}(1)\, \rho_1^{\Wn}(2) + 
 \rho_1^{\Wp}(2)\, \rho_1^{\Wn}(1)\right\}    
\label{eq:rho2c}
\end{eqnarray}
Defining the experimentally  often studied {\em normalized\/}  two-particle density 
\begin{equation}
R_2^{\WW}(1,2)=
\frac{\rho_2^{\WW}(1,2)}{\rho_1^{\WW}(1)\,\rho_1^{\WW}(2)},
\label{eq:ratio:def}
\end{equation}
one  finds
\begin{equation}
R_2^{\WW}(1,2)=\frac{%
 \rho_2^{\Wp}(1,2) + \rho_2^{\Wn}(1,2)+
\left\{ \rho_1^{\Wp}(1)\, \rho_1^{\Wn}(2) + 
        \rho_1^{\Wp}(2)\, \rho_1^{\Wn}(1)\right\}\,
\left\{  1+\delta_I(1,2) \right\}}{
\rho_1^{\Wp}(1)\, \rho_1^{\Wp}(2) + 
\rho_1^{\Wn}(1)\, \rho_1^{\Wn}(2) +
\rho_1^{\Wp}(1)\, \rho_1^{\Wn}(2) + 
\rho_1^{\Wp}(2)\, \rho_1^{\Wn}(1)
}.%
\label{eq:ratio:a} 
\end{equation}

\subsection{Correlations in the variable $Q$}
The previous sections dealt exclusively with fully differential quantities.
In practice, these are impossible to measure 
and a projection on a lower-dimensional space is needed.
We here consider, for illustration, the kinematical variable  $Q^2=-(p_1-p_2)^2$, 
the negative square of the difference in 4-momenta of particles 1 and 2.
We  use the notation 
$\rho^{\Wp}\otimes \rho^{\Wn}(Q)$
 for  integrals of the type
\begin{equation}
\int\int d^3p_1d^3p_2\, \rho^{\Wp}(1)\,\rho^{\Wn}(2)\,\delta\left(Q^2+(p_1-p_2)^2\right).
\label{eq:cross}
\end{equation}
In practical applications, such integrals are  calculated 
using an event and track-mixing technique.

We assume from now on that  $\delta_I(1,2)$ in (\ref{eq:cumulww}) is a function of $Q$ only: 
$\delta_I(Q)$.
This  simplifies the calculations but may not be
fully realistic. At least for Bose-Einstein   studies, it is known that
the correlation function of like-sign pairs is not isotropic in four-momentum space.
To simplify further we also assume that 
$$\rho_1^{\Wp}\otimes \rho_1^{\Wp}(Q)=\rho_1^{\Wn}\otimes \rho_1^{\Wn}(Q)
\equiv \rho_1^{\W}\otimes \rho_1^{\W}(Q).$$

The expressions in the previous sections take  the following form
\begin{eqnarray}
\rho_2^{\WW}(Q)&=&\rho_2^{\Wp}(Q)+ \rho_2^{\Wn}(Q)+  
2\left\{1+\delta_I(Q)\right\} \,\rho_1^{\Wp}\otimes\rho_1^{\Wn}(Q), \label{eq:q:1}\\
C_2^{\WW}(Q)&=&C_2^{\Wp}(Q)+ C_2^{\Wn}(Q)+ 
2\delta_I(Q)\,\left\{ \rho_1^{\Wp}\otimes \rho_1^{\Wn}(Q)\right\}.
\label{eq:q:2}
\end{eqnarray} 
Integrating these expressions over all $Q$, we  have
\begin{equation}
 \delta_I=\frac{1}{\aver{n_{W^+}}\,\aver{n_{W^-}} }
\int dQ\,\delta_I(Q)\,
\left\{ \rho_1^{\Wp}\otimes \rho_1^{\Wn}(Q)\right\}.
\label{eq:integrated:delta} 
\end{equation}
Further
\begin{equation}
R_2^{\WW}(Q)=
\frac{%
\rho_2^{\Wp}(Q)+\rho_2^{\Wn}(Q) +  
2\left\{1+\delta_I(Q)\right\} \,\left\{ \rho_1^{\Wp}\otimes \rho_1^{\Wn}(Q)\right\} }
{  2\left\{ \rho_1^{\W}\otimes \rho_1^{\W}(Q) + \rho_1^{\Wp}\otimes \rho_1^{\Wn}(Q)\right\}
}.
\label{eq:q:3}
\end{equation}

Introducing  the ``overlap function''
\begin{equation}
g(Q)=\frac{\rho_1^{\Wp}\otimes \rho_1^{\Wn}(Q)}{\rho_1^{\W}\otimes \rho_1^{\W}(Q)},
\label{eq:overlapfunction} 
\end{equation}
we can rewrite (\ref{eq:q:1})-(\ref{eq:q:2}) as
\begin{eqnarray}
\rho_2^{\WW}(Q)&=&\rho_2^{\Wp}(Q)+ \rho_2^{\Wn}(Q)+  
2\left\{1+\delta_I(Q)\right\}\,g(Q)\,\rho_1^{\W}\otimes\rho_1^{\W}(Q)
\label{eq:q:4},\\
C_2^{\WW}(Q)&=&C_2^{\Wp}(Q)+ C_2^{\Wn}(Q)+ 
2\delta_I(Q)\,g(Q)\,\rho_1^{\W}\otimes \rho_1^{\W}(Q).\label{eq:q:5}
\end{eqnarray}
Likewise
\begin{eqnarray}
R_2^{\WW}(Q)&=&
\frac{%
\rho_2^{\Wp}(Q)+\rho_2^{\Wn}(Q) +  2\left\{1+\delta_I(Q)\right\}\,g(Q) \,
 \rho_1^{\W}\otimes \rho_1^{\W}(Q) }
{  2\rho_1^{\W}\otimes \rho_1^{\W}(Q)\left\{ 1+g(Q) \right\}
},
\label{eq:q:6}\\
&=&\frac{1}{2} \frac{R_2^{\Wp}(Q)}{1+g(Q)}+
\frac{1}{2}    \frac{R_2^{\Wn}(Q)}{1+g(Q)}+
\left\{ 1+\delta_I(Q)\right\} \frac{g(Q)}{1+g(Q)}.\label{eq:q:7}
\end{eqnarray} 
Here $R_2^{\W}(Q)$  is the normalized two-particle density for a single $W$
\begin{equation}
R_2^{\W}(Q)=\frac{\rho_2^{\W}(Q)}{\rho_1^{\W}\otimes\rho_1^{\W}(Q)}=1+K_2^{\W}(Q),
\label{eq:def:r2single} 
\end{equation}
with  $K_2^{\W}(Q)$  the normalized two-particle cumulant density.

The meaning of the various terms in (\ref{eq:q:4}) is the following:
a term such as $\rho_2^{\Wp}(Q)$ ``counts'' the number of like-sign pairs within a single
$W^+$. 
Integrated over all $Q$, it equals $\aver{n_{\Wp}(n_{\Wp}-1)}$;
the term $2\left\{1+\delta_I(Q)\right\}g(Q)\rho_1^{\W}\otimes\rho_1^{\W}(Q)$, ``counts''
the number of pairs ($i,j$) with particle $i$ belonging to $W^+$ and particle $j$ belonging
to $W^-$. Its integral over all $Q$ is equal to $2\aver{n_{\Wp}n_{\Wn}}$;
 the term $2g(Q)\rho_1^{\W}\otimes\rho_1^{\W}(Q)\equiv \rho_1^{\Wp}\otimes\rho_1^{\Wn}(Q)$ 
``counts'' the number of pairs ($i,j$) in uncorrelated $W$'s. Its integral equals
$2\aver{n_{\Wp}}\aver{n_{\Wn}}$. These relations can serve as a check of any experimental method
used to calculate the so-called ``mixing'' terms $\rho_1^{\W}\otimes\rho_1^{\W}(Q)$, 
$\rho_1^{\Wp}\otimes\rho_1^{\Wn}(Q)$. 

To simplify further the expressions, but without essential loss of generality, 
 we shall from now on assume that
\begin{equation}
\rho_2^{\Wp}(Q)=\rho_2^{\Wn}(Q)\equiv\rho_2^{\W}(Q).
\label{eq:rho2:equal} 
\end{equation}
We obtain from (\ref{eq:q:7})
\begin{eqnarray}
R_2^{\WW}(Q)&=&  \frac{R_2^{\W}(Q)}{1+g(Q)}+
\left\{1+\delta_I(Q)\right\}\,\frac{g(Q)}{1+g(Q)}=1+
\frac{K_2^{\W}(Q)+g(Q)\delta_I(Q)}{1+g(Q)},
\label{eq:q:7a}\\
K_2^{\WW}(Q)&=&\frac{K_2^{\W}(Q)}{1+g(Q)} +\delta_I(Q)\frac{g(Q)}{1+g(Q)}.
\label{eq:q:7b} 
\end{eqnarray} 
where $K_2^{\WW}(Q)=R_2^{\WW}(Q)-1$, is the normalized  cumulant density
of the $WW$ system.

Both  $K_2^{\WW}(Q)$ and $K_2^{\W}(Q)$
  are often parameterized with  Gaussians.
However, from Eqs.~(\ref{eq:q:7a})-(\ref{eq:q:7b}), it is seen that neither
$R_2^{\WW}(Q)$ nor $K_2^{\WW}(Q)$ will, in general, have 
the same functional $Q$-dependence as  $K_2^{\W}(Q)$ unless $K_2^{\W}(Q)\sim\delta_I(Q)$ and
$g(Q)$ is constant.

Consider next the following limiting forms of the previous expressions.
\begin{itemize}
\item {\bf Fully  overlapping and uncorrelated \boldmath{$W$}-decays: \boldmath{$\delta_I(Q)=0$}.}

Here, all factors $\rho_1\otimes \rho_1(Q)$ are equal so that $g(Q)=1$.  
Equations~(\ref{eq:q:7a}-\ref{eq:q:7b}) become
\begin{eqnarray}
R_2^{\WW}(Q)&=&
\frac{1}{2}\, \left\{1+R_2^{\W}(Q)\right\},
\label{eq:r2ww}\\
K_2^{\WW}(Q)&=&
\frac{1}{2}\, K_2^{W}(Q).
\label{eq:k2ww} 
\end{eqnarray}
For this special case, the cumulant correlation function  for  $W^+W^-\to4q$ events is 
only half  that of a  single $W$, corresponding
to $W^+W^-\to2q$ events. This result was first derived in~\cite{edw:beww}.
Its validity basically rests on the additivity of factorial cumulants  for sums of
independent random variables%
%
\footnote{In general, 
for $S$ {\em fully overlapping identical and independent\/} ``sources'' it follows from the additivity of
cumulants that
\begin{equation}
\frac{\tilde{K}^{(S)}_q}{\aver{n}^q}=\frac{\tilde{K}^{(1)}_q}{S^{q-1}},
\label{eq:cumulant:dilution} 
\end{equation}
with $\tilde{K}^{(S)}_q$ the unnormalized 
integrated factorial cumulant of a system composed of
$S$ sources, $\aver{n}$ its average multiplicity
 and $\tilde{K}^{(1)}_q$ the integrated factorial 
cumulant of a single source~\cite{lipa:bush}.
We note that, besides the assumptions mentioned, no further approximations are
involved in deriving~(\ref{eq:cumulant:dilution}) in contrast to the 
derivation in~\cite{gideon:sources} where (\ref{eq:cumulant:dilution}) is only obtained
under (unnecessary) further conditions.}.

\item  {\bf Decreasing overlap, arbitrary \boldmath{$\delta_I(Q)$}.}
Above production threshold, and with increasing center-of-mass energy, $\sqrt{s}$,
 the decay products of the two $W$'s will show a decreasing overlap in momentum space.
The factor $\rho_1^{\Wp}\otimes\rho_1^{\Wn}(Q)$  
which is a measure of this
overlap, will  tend to zero  for any fixed value of $Q$ when $s\to\infty$.
For $g(Q)\to 0$ one obtains
\begin{eqnarray}
R_2^{\WW}(Q)&\rightarrow&\frac{%
\rho_2^{\W}(Q)
}%
{%
\rho_1^{\W}\otimes\rho_1^{\W}(Q)
}=R_2^{\W}(Q),
\label{eq:r2:nooverlap2}\\
K_2^{\WW}(Q)&\rightarrow&
K_2^{W}(Q).
\label{eq:r2:nooverlap} 
\end{eqnarray}
This result is intuitively clear. The diminishing  overlap of the
$W^+$ and $W^-$  decay products  decreases the contribution from 
pairs of  particles which 
originate from different $W$'s. For any fixed  $Q$, 
$R_2^{\WW}(Q)$ increasingly receives  contributions  
from pairs belonging to the same
$W$ only. 
\end{itemize}

Equations (\ref{eq:q:7a}-\ref{eq:q:7b}) 
show that, as a consequence of not fully overlapping ``sources'',  
 genuine inter-source correlations, 
represented by $\delta_I(Q)$, are ``diluted''  in the actual measurement
of $R_2^{\WW}(Q)$  or $K_2^{\WW}(Q)$ since $g(Q)/(1+g(Q))$
 may become quite small in practical situations.

This has further implications. 
In experimental conditions where  one is attempting to
avoid the effect of inter-source correlations (such as in $W$-mass measurements), 
methods should be
devised to render the overlap function
as small as possible {\eg} by suitably chosen cuts.

On the opposite, when the main interest is focussed on establishing 
inter-source correlations, one could try to devise methods that 
maximize  $g(Q)/(1+g(Q))$ and thus increase  the experimental sensitivity to 
$\delta_I(Q)$.

With respect to searches for inter-$W$ correlations at LEP,
it remains to be investigated  how much the presently adopted cuts and 
algorithms used to select  with high efficiency $WW$ events, 
affect the shape of the overlap fuction and contribute to  
weaken  (possible) genuine $WW$ correlation effects in the analyzed data.
The same holds for studies of possible color-reconnection effects.

With the formalism described above, it is now straightforward to obtain expressions
for a variety of observables used in experimental studies of
inter-$W$ correlations. Some of these are reviewed in the following. Others are easily
constructed using the results of this section.

\section{Examples}
\subsection{The L3 test-statistics \boldmath{$D$} and \boldmath{$D'$}}
The L3 collaboration, in a search for inter-$W$ (BE) correlations,
discussed in a recent paper~\cite{l3:wwbe}
the distributions (``test-statistics'')
\begin{eqnarray}
\Delta\rho(Q)&=&\rho_2^{\WW}(Q) -2\rho_2^{\W}(Q)-2\rhomix(Q) \label{eq:q:delta},\\
D(Q) &=& \frac{\rho_2^{\WW}(Q)}{2\rho_2^{\W}(Q)+2\rhomix(Q)} \label{eq:q:d},\\
D'(Q)&=&\frac{ D(Q) }{D_{\mbox{\rm\small MC, no BEC}}(Q)}; \label{eq:q:dprime}
\end{eqnarray}
where $\rhomix(Q)$ is identical to $\rho_1^{\Wp}\otimes \rho_1^{\Wn}(Q)$.
The ratio $D'(Q)$ is obtained by dividing $D(Q)$ by the same derived
 from a Monte Carlo calculation {\em without\/} Bose-Einstein correlations.
As very similar quantity was studied by ALEPH~\cite{aleph:wwbe}.

These distributions can  be rewritten as
\begin{eqnarray}
\Delta\rho(Q)&=&
2g(Q)\,\delta_I(Q)\,\rho_1^{\W}\otimes\rho_1^{\W}(Q),  \label{eq:q:delta1}\\
D(Q) &=& 1+ \delta_I(Q)\frac{g(Q)}{R_2^{\W}(Q)+g(Q)},\label{eq:q:d1}\\
D'(Q)&=& 1+ \delta_I(Q)\frac{g(Q)}{R_2^{\W}(Q)+g(Q)}.
\label{eq:q:dprime1}
\end{eqnarray}
Observe that (\ref{eq:q:d1}) and (\ref{eq:q:dprime1}) are formally, but not necessarily 
experimentally, identical\footnote{Eq.(\ref{eq:q:dprime1}) remains unchanged when it is calculated 
from a Monte Carlo with only intra-$W$ correlations.}.

These expressions show that 
none of the studied observables isolates
completely the genuine inter-$W$ correlation function $\delta_I(Q)$. 
It can be measured most directly
via the unnormalized function $\Delta\rho(Q)$, as seen from (\ref{eq:q:delta1}), once 
the mixing term $\rho_1^{\Wp}\otimes\rho_1^{\Wn}(Q)$ is determined.
This equation therefore (strongly) suggests to study the  ratio
\begin{equation}
\frac{\Delta\rho(Q)}{2\rhomix(Q)}=\delta_I(Q).
\label{eq:deltarho:normalized} 
\end{equation}
It can be determined directly from data only.

\subsection{The fraction of pairs from different \boldmath{$W$}'s}
This distribution  was used by
L3~\cite{vandalen}  and  DELPHI~\cite{delphi:wwbe:osaka}.
The fraction of like sign pairs from {\em different\/}  $W$'s is, in our notation, defined as
\begin{equation}
F(Q)=\frac{2\left\{1+\delta_I(Q)\right\}\,\rho_1^{\Wp}\otimes \rho_1^{\Wn}(Q)}{\rho_2^{\WW}(Q)},
\label{eq:fraction1} 
\end{equation}

or
\begin{equation}
F(Q)=\frac{g(Q)\left\{ 1+\delta_I(Q)\right\} }{R_2^{\W}(Q)+g(Q)\left\{1+\delta_I(Q)\right\}}.
\label{eq:fraction2} 
\end{equation}
The equation can be solved for $g(Q)$ with the result
\begin{equation}
g(Q)=R_2^{\W}(Q)\, \frac{F(Q)}{1-F(Q)}\,\frac{1}{1+\delta_I(Q)}.
\label{ccc} 
\end{equation}
Using $F(Q)$ presented in~\cite{vandalen}, we have plotted  $g(Q)$ and $g(Q)/(1+g(Q))$ 
in Fig.~1. Since $\delta_I(Q)$ is 
unknown, we assume, for illustration,  the ``Goldhaber''-form
$\delta_I(Q)=\Lambda \exp{(-r_I^2Q^2)}$ and $R_2^{\W}=1+\lambda\exp{(-r^2Q^2)}$, with
$r=0.67$~fm and $\lambda=0.7$ taken from~\cite{l3:wwbe}. We set $r_I=r$.
 
The function $g(Q)/(1+g(Q))$ has a value between  $0.14-0.24$ at $Q=0$, 
depending on $\Lambda$, implying ({\em cfr.\xspace} Eq.~(\ref{eq:q:7b}))
that the effectively measured strength of possible inter-$W$ correlation effects is smaller 
 by the same factor.

With $g(Q)$ derived from the L3 results, we calculated the
quantity $D(Q)$ as a function of $\Lambda$. The result is shown in Fig.~2.
As is clear from (\ref{eq:q:d1}), $D(Q)$ {\em is not\/} of the form $\sim(1+\tilde{\Lambda}
\exp{(-k^2Q^2)})$,  used in~\cite{l3:wwbe}. 

To test for inter-$W$ correlations, a density function, $R_2^{\WW}(Q)_{\mbox{\rm\tiny BES}}$,
 is defined in~\cite{vandalen}  which is supposed 
to coincide with the data distribution $R_{2,L3}^{\WW}(Q)$ (defined below)  
if  inter-$W$ correlations are absent. 
This function is  stated to describe the data on $R_{2,L3}^{\WW}(Q)$, 
thus adding extra  support to the claimed absence of inter-$W$ BEC.

The quantity $R_{2,L3}^{\WW}(Q)$ is
defined as the ratio of the like-sign two-particle density in the data 
to the same obtained  from a
Monte Carlo model without Bose-Einstein correlations. 
Assuming that the  model  reproduces
perfectly $R_2^{\W}(Q)$, this quantity is given by
\begin{equation}
R_{2,L3}^{\WW}(Q)=\frac{R_2^{\W}+g(Q)\left\{1+\delta_I(Q)\right\}}{1+g(Q)}=1+
\frac{K_2^{\W}+g(Q)\delta_I(Q)}{1+g(Q)}.\label{eq:q:11} 
\end{equation}
The ``test'' function  $R_2^{\WW}(Q)_{\mbox{\rm\small\tiny BES}}$ is taken to be
\begin{equation}
R_2^{\WW}(Q)_{\mbox{\rm\tiny BES}}= 1+ \left\{1-F(Q)\right\}\, K_2^{\W}(Q),
\label{eq:vandalen} 
\end{equation}
with  $F(Q)$  given by  (\ref{eq:fraction2}) and setting  $\delta_I(Q)=0$.
Equation~(\ref{eq:vandalen}) can then be written as
\begin{equation}
R_2^{\WW}(Q)_{\mbox{\rm\small \tiny BES}}= 1+ 
\frac{1+K_2^{\W}(Q)}{1+K_2^{\W}(Q)+g(Q)}\, K_2^{\W}(Q).
\label{eq:vandalen2} 
\end{equation}
With $\delta_I(Q)=0$ in (\ref{eq:q:11}) for no inter-$W$ correlations, 
it is seen that this distribution and that of 
Eq.~(\ref{eq:vandalen2}) are different, unless $K_2^{\W}(Q)=0$  or $g(Q)=0$. 
We conclude that 
(\ref{eq:vandalen}) is not suitable to test for the ``null hypothesis'' $\delta_I(Q)=0$.
An analogous test-distribution,  used by the DELPHI Collaboration for similar purposes 
(see below), has the same defect.

\subsection{The DELPHI observables}
In the DELPHI analysis of Bose-Einstein 
correlations in $WW$ events~\cite{delphi:wwbe:osaka}, the following ratio is
studied
\begin{equation}
R_{\mbox{\rm\small 4q}}^{\mbox{\rm\small data}}=
\frac{\rho_2^{\WW}(Q)}{\rho_{\mbox{\rm\small 2,MC, no BE}}^{\WW}(Q)},
\label{eq:delphi:1} 
\end{equation}
where the denominator is calculated from a Monte Carlo without Bose-Einstein effects.
This can be rewritten using (\ref{eq:q:4}) as
\begin{equation}
R_{\mbox{\rm\small 4q}}^{\mbox{\rm\small  data}}(Q)=\frac{R_2^{\W}(Q)}{1+g(Q)}
+\left\{1+\delta_I(Q)\right\}\frac{g(Q)}{1+g(Q)}
=1+\frac{K_2^{\W}(Q)+g(Q)\delta_I(Q)}{1+g(Q)},
\label{eq:delphi:2} 
\end{equation}
which coincides with the general expression (\ref{eq:q:7a}).
We have assumed that, in the Monte Carlo, identical particle pairs are uncorrelated, 
i.e.~$\rho_2^{\W}(Q)=\rho_1^{\W}\otimes\rho_1^{\W}(Q)$.

The function (\ref{eq:delphi:2}) is plotted in Fig.~\ref{fig:delphi:r}, together with
$R_2^{\W}$ (denoted by $R_{\mbox{\rm\small 2q}}$ in~\cite{delphi:wwbe:osaka}) measured in
 the DELPHI analysis and  used here as input.
From the above expression, and from the figure, it is clear that this distribution  remains
sensitive to inter-$W$ correlations. However, it cannot, in general,  be described
by the same parameterization as used for  $R_2^{\W}(Q)$, even for $\delta_I(Q)=0$.
Fig.\ref{fig:delphi:r} also shows, for illustration, the same function  
using a Gaussian parameterization of $\delta_I(Q)$  with a ``radius-parameter'' of 3~fm.
The latter might be typical  for the range of second-order interference (HBT) correlations
associated with incoherent strings. In this case, the $Q$-region affected by BEC is
restricted to values below $\sim100$~MeV and will decrease even further for still 
larger ``radii''. 

In Fig.~\ref{fig:delphi:intercept} we show the intercept,
$R_{\mbox{\rm\small 4q}}^{\mbox{\rm\small  data}}(Q)$ at $Q=0$, and that of $R_2^{\W}(Q)$, as a function
of $\Lambda$.  It is seen that the intercept for the $WW\to 4q$ channel remains below that
of   $R_2^{\W}(Q)$ for $\Lambda<\lambda$. It is, of course, independent of the
values of the parameters $r$ and $r_I$.

\subsubsection{\boldmath{$R_{\mbox{\rm\small 4q}}(Q)(\mbox{\rm\small \bf mixing})$}}
DELPHI also  studied a ``test'' distribution, 
$R_{\mbox{\rm\small 4q}}(Q)(\mbox{\rm\small mixing})$, 
constructed from  mixed  independent ($WW\to 2q$) events and defined as
\begin{equation}
 R_{\mbox{\rm\small 4q}}(Q)(\mbox{\rm\small mixing})=\frac{%
\left[\rho_2^{\W}(Q)+\rho_2^{\mbox{\rm\small mix}}(Q)\right]_{\mbox{\rm\small data}}}{%
\left[\rho_2^{\W}(Q)+\rho_2^{\mbox{\rm\small mix}}(Q)\right]_{\mbox{\rm\small MoCa, No Be}}}
\label{eq:delphi:3} 
\end{equation}
with $ \rho_2^{\mbox{\rm\small mix}}(Q)=\rho_1^{\Wp}\otimes\rho_1^{\Wn}$.
The above expression is equal to
\begin{equation}
 R_{\mbox{\rm\small 4q}}(Q)(\mbox{\rm\small mixing})=1+
\frac{K_2^{\W}(Q)}{1+g(Q)}.
\label{eq:delphi:4} 
\end{equation}
This expression indeed coincides with (\ref{eq:delphi:2}) for $\delta_I(Q)=0$
and can thus be used to test for inter-$W$ correlations. 
It is plotted in Fig.~\ref{fig:delphi:r} and depends here (weakly) on
$\delta_I(Q)$ which enters the definition of $g(Q)$.
We remark again that $R_{\mbox{\rm\small 4q}}(Q)(\mbox{\rm\small mixing})$  
cannot, in general,  be described
by the same parameterization as used for  $R_2^{\W}(Q)$.

Defining $\Delta\lambda(\mbox{\rm\small mixing})$ as the difference in intercepts
at $Q=0$ of   $R_{\mbox{\rm\small 4q}}(Q)(\mbox{\rm\small mixing})$ and 
$R_{\mbox{\rm\small 4q}}^{\mbox{\rm\small data}}$
we find
\begin{equation}
\Delta\lambda(\mbox{\rm\small mixing})= \Lambda\, \frac{g(0)}{1+g(0)},
\label{eq:delphi:5} 
\end{equation}
with  $\Lambda=\delta_I(0)$.

\subsubsection{\boldmath{$R_{\mbox{\rm\small 4q}}(Q)(\mbox{linear})$}}
A further quantity studied by DELPHI is  $R_{\mbox{\rm\small 4q}}(Q)(\mbox{linear})$, defined in
Eq.~(25) of ref.~\cite{delphi:wwbe:osaka}. It is constructed to supposedly
coincide with $R_2^{\WW}(Q)$ if inter-$W$ correlations are absent. 
With our notation it is given by
\begin{equation}
R_{\mbox{\rm\small 4q}}(Q)(\mbox{linear})=
R_2^{\W}(Q)+\left\{ 1-R_2^{\W}(Q)\right\}\,\frac{g(Q)}{R_2^{\W}(Q)+g(Q)}.
\label{eq:delphi:xx} 
\end{equation}
This function is  plotted in Fig.~\ref{fig:delphi:r2}.
The distribution differs  from $R_{\mbox{\rm\small 4q}}^{\mbox{\rm\small data}}$  
even for $\delta_I(Q)=0$ (or $\Lambda=0$) and, therefore, has not the intended properties.
Nevertheless, from Fig.~3 and Fig.~5 
it can be seen that $R_{\mbox{\rm\small 4q}}(Q)(\mbox{linear})$ and
$R_{\mbox{\rm\small 4q}}(Q)(\mbox{mixing})$ are numerically quite close. 
The weak dependence on $\Lambda$ is again due to the apriori unknown  dependence of $g(Q)$ on $\delta_I(Q)$.

\section{Summary}
In this paper we described a mathematical formalism which should allow a systematic
study and improved understanding of particle correlations in a physical system which is
composed of $S$ possibly stochastically correlated parts.
Most emphasis was put on two-particle inclusive densities and correlation functions
for $S=2$, but the
formalism can be extended to arbitrary orders and arbitrary $S$.

Taking the reaction 
$e^+e^-\to W^+W^-\to q_1\overline{q}_2 q_3\overline{q}_4\to \mbox{hadrons}$ as
a prime example, it was shown how to relate the measurement of
correlations within one hadronically decaying $W$ 
to those measured in the full $WW\to 4q$ system.

For  the observables used in present  studies, 
correlations in  fully hadronic $WW$ decay's
can be expressed in terms of the correlation function of a single $W$, 
an inter-$W$ correlation
$\delta_I$ and a function, $g$, 
which quantifies the degree of overlap in momentum space of the decay products from different $W$'s. 

On several examples, we have shown that some  of the presently used techniques to
search for inter-$W$ correlations are not optimal. 
In some cases, experimental quantities have been
used which lack the correct mathematical properties and are  based on
intuition rather than on sound mathematics.

The determination of the overlap function 
allows to assess quantitatively the sensitivity of a particular
distribution to inter-$W$ correlations. Moreover, the influence of 
experimental and methodological cuts  can be systematically investigated . 

An estimate of $g(Q)$, presented here, indicates that the fully hadronic, four-jet
decay channel of the reaction $e^+e^-\to W^+W^-$ may not be optimal to search for
inter-$W$ Bose-Einstein effects. The usual event selections, requiring four well-isolated
jets in the final state, weaken the sensitivity to possible inter-$W$ correlations.
Better results may well be obtained in analyses using less stringent cuts, 
or which include  {\eg} three-jet events  with suitably chosen topology so as to keep the
QCD background at a tolerable level.

Our survey of some of the  test-distributions used so far indicates that the
observable $\Delta\rho(Q)$, defined in Eq.~(\ref{eq:q:delta}),
proposed in~\cite{edw:beww} and measured by L3~\cite{l3:wwbe},
provides a direct measure of possible inter-$W$ correlations.
It suggests itself as suitable observable well-adapted
for a global LEP-wide combination of $WW$ data.

\begin{figure}[ht]
\begin{center}
\epsfig{file=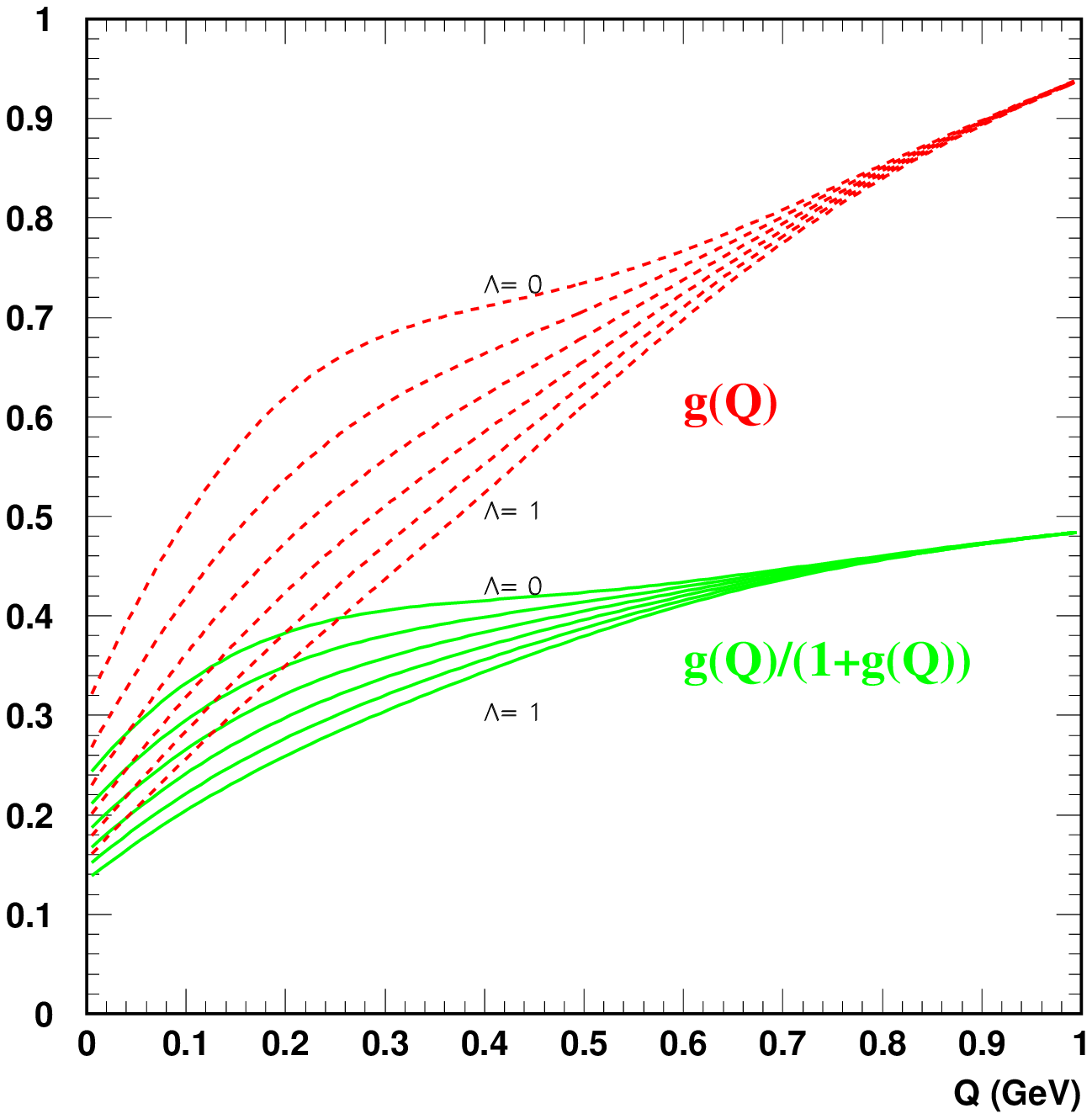}
\caption{L3: the functions $g(Q)$ and  $g(Q)/(1+g(Q))$;
$g(Q)$ is calculated assuming the form
$\delta_I(Q)=\Lambda\exp{(-r^2_IQ^2)}$, $r_I=0.67$~fm,
 with  $\Lambda$ varying in the range $0.0-1.0$, in steps of $0.2$.}
\label{fig:l3:g}
\end{center}

\end{figure}

\begin{figure}[ht]
\begin{center}
\epsfig{file=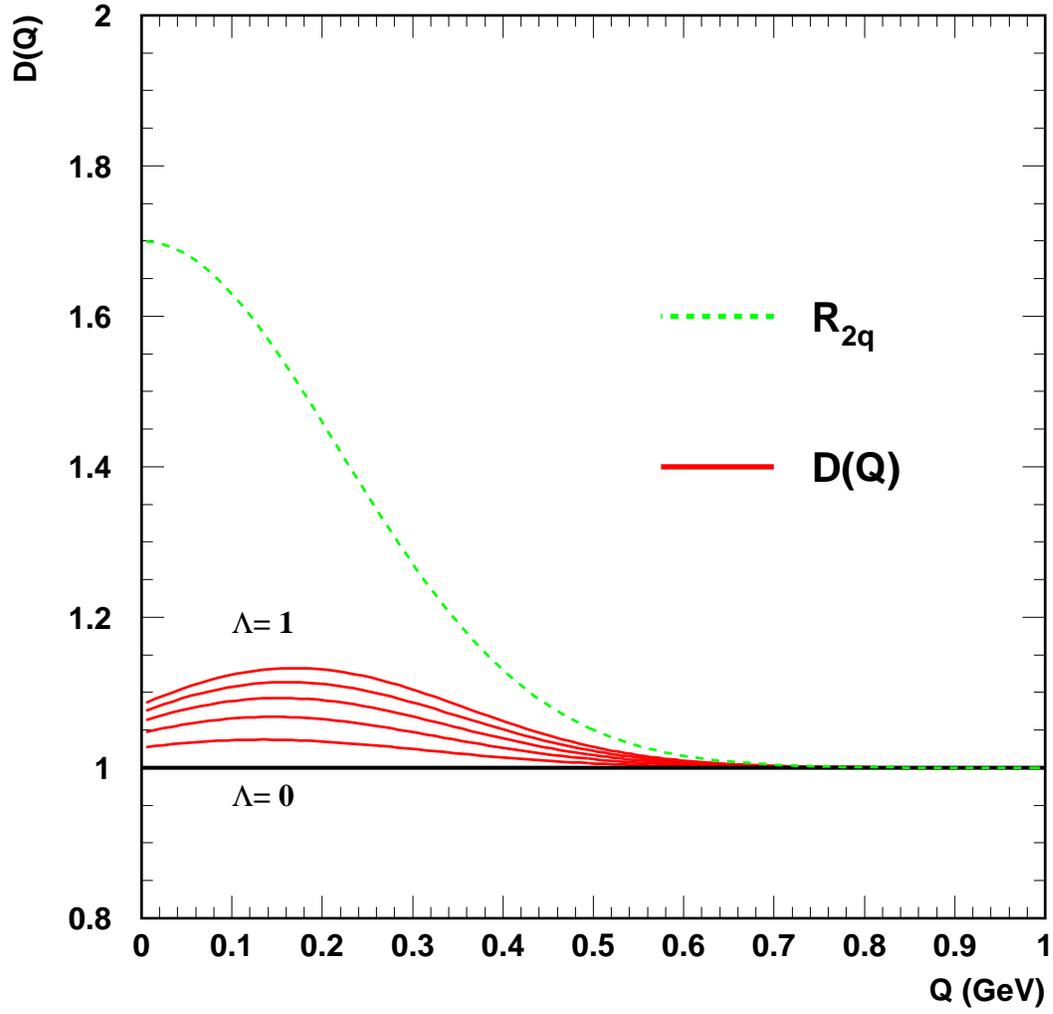}
\caption{The test-statistic $D(Q)$ used by L3, Eq.(\ref{eq:q:d1}). Also shown is the L3 
parameterization  of $R_{2q}(Q)\equiv R_2^{\W}(Q)$.}\label{fig:l3:d}
\end{center}

\end{figure}

\begin{figure}[ht]
\begin{center}
\epsfig{file=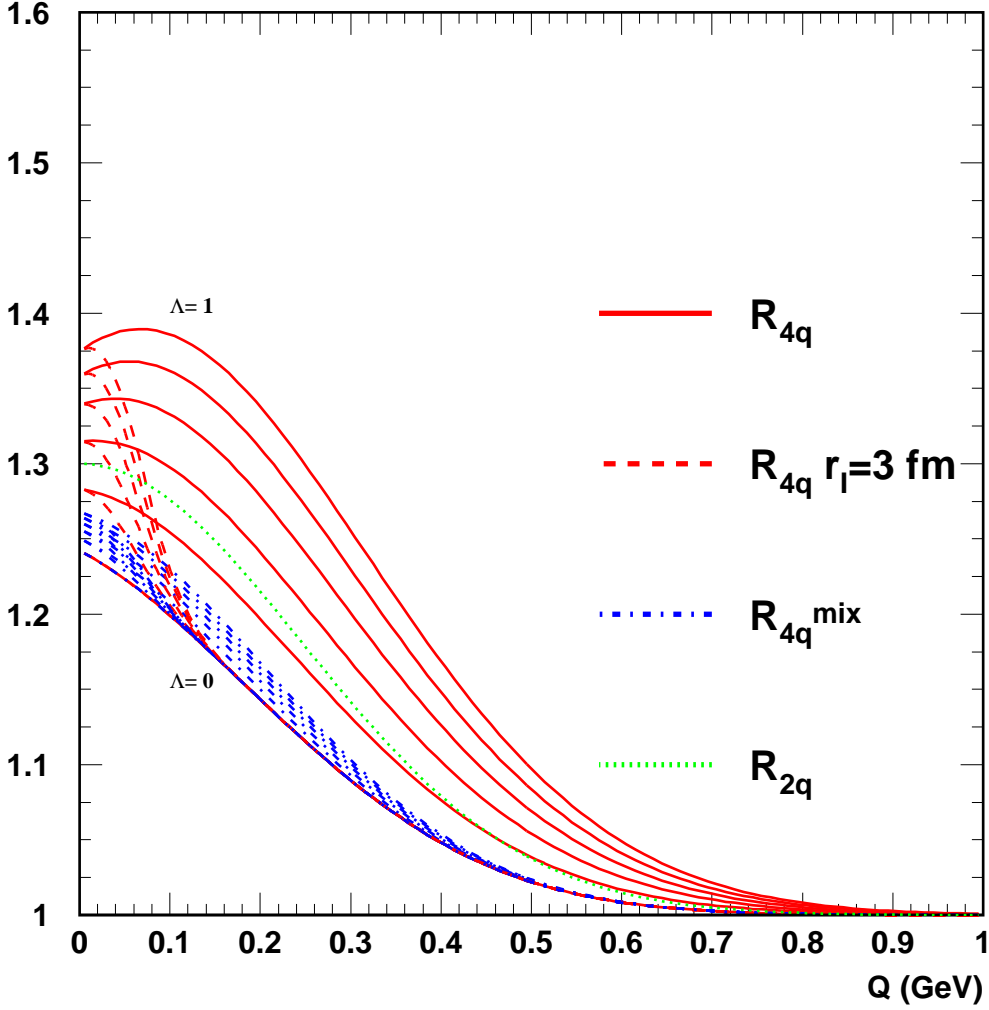}
\caption{DELPHI: The normalized densities $R_{4q}(Q)=R_2^{\WW}(Q)$, 
Eq.(\ref{eq:delphi:2}) in the text, using as input the DELPHI parameterization
of  $R_{2q}(Q)\equiv R_2^{\W}(Q)$ and assuming 
$\delta_I(Q)=\Lambda\exp{(-r^2Q^2)}$; $\Lambda$ varies in the range $0.0-1.0$, in steps of $0.2$
The dashed curves show the same with $r=3$~fm. Also shown (dot-dashed)  is the distribution,
Eq.(\ref{eq:delphi:4}), corresponding to mixed independent $WW\to 2q$ events. It depends
implicitly  on $\delta_I(Q)$ through the function $g(Q)$.}
\label{fig:delphi:r}
\end{center}
\end{figure}

\begin{figure}[ht]
\begin{center}
\epsfig{file=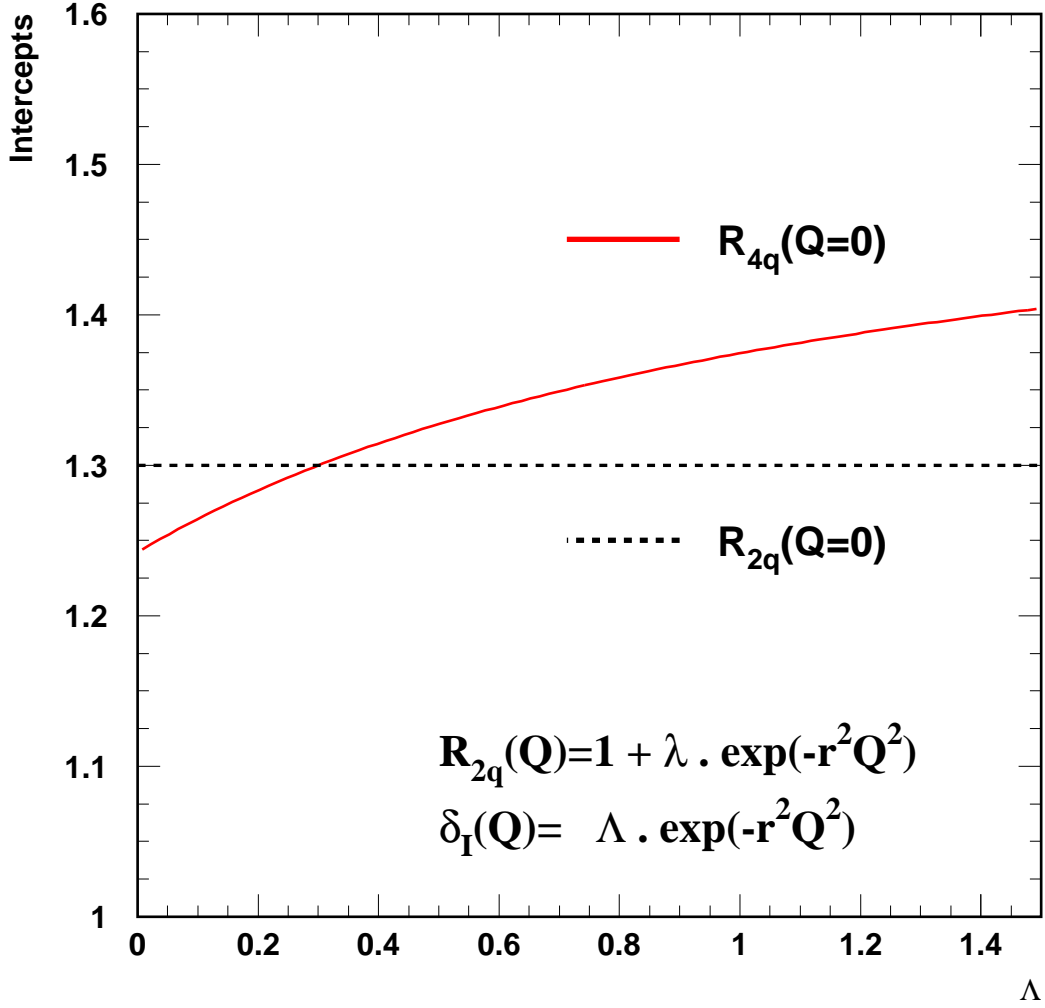}
\caption{DELPHI: The 
value of the normalized density $R_{4q}(Q)\equiv R_2^{\WW}(Q)$, 
Eq.(\ref{eq:delphi:2}) in the text, 
and of $R_{2q}(Q)\equiv R_2^{\W}(Q)$, at $Q=0$, as a function of the parameter $\Lambda$ in
$\delta_I(Q)=\Lambda\exp{(-r^2Q^2)}$.}
\label{fig:delphi:intercept}
\end{center}

\end{figure}

\begin{figure}[ht]
\begin{center}
\epsfig{file=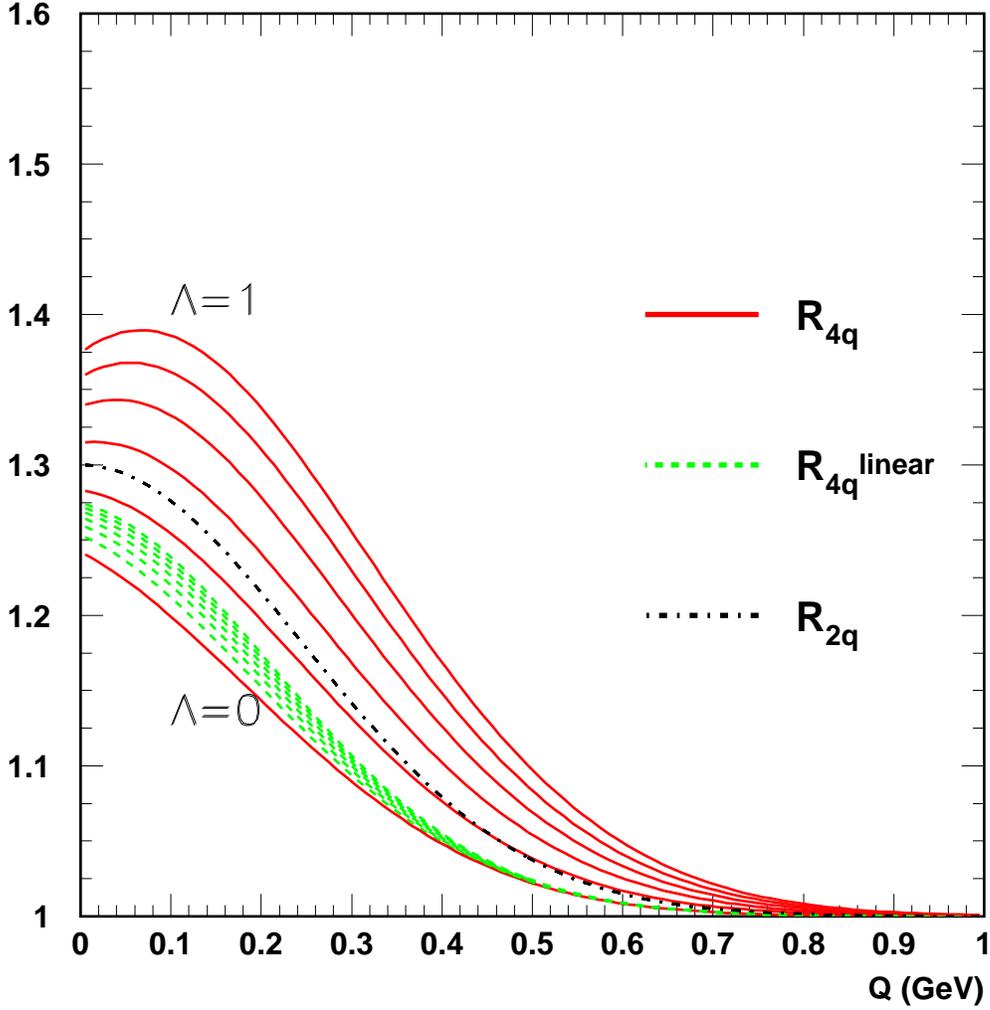}
\caption{DELPHI: The normalized density $R_{4q}(Q)\equiv R_2^{\WW}(Q)$, 
Eq.(\ref{eq:delphi:2}) in the text, using as input the DELPHI parameterization of
 $R_{2q}(Q)\equiv R_2^{\W}(Q)$ and with 
$\delta_I(Q)=\Lambda\exp{(-r^2Q^2)}$; $\Lambda$ varies in the range $0.0-1.0$ in steps of $0.2$.
Also shown is the distribution corresponding to the so-called 
Linear Scenario in ref.~\cite{delphi:wwbe:osaka}.
\label{fig:delphi:r2}}
\end{center}
\end{figure}


\end{document}